\documentclass[authoryear, preprint,12pt]{elsarticle}
\usepackage{amsmath}

\usepackage{amssymb}

\usepackage[colorlinks=true,citecolor=blue, linkcolor=blue]{hyperref}
\usepackage[normalem]{ulem}
\usepackage{subcaption}

\usepackage{xcolor}
\usepackage{float}
\usepackage{appendix}
\journal{arXiv}
\begin{document}

\begin{frontmatter}

\title{Estimating cross-border mobility from the difference in peak-timing: A case study in Poland-Germany border regions }

\author[inst1,inst2]{Abhishek Senapati\corref{corauthor}}
\author[inst1,inst2]{Adam Mertel}
\author[inst1,inst2]{Weronika Schlechte-Welnicz}
\author[inst1,inst2,inst3,inst4]{Justin M. Calabrese}
\cortext[corauthor]{Corresponding author\\ Email: abhishekiz4u04@gmail.com}

\address[inst1]{Center for Advanced Systems Understanding (CASUS), Goerlitz, Germany}
\address[inst2]{Helmholtz-Zentrum Dresden-Rossendorf (HZDR), Dresden, Germany}
\address[inst3]{Helmholtz Centre for Environmental Research-UFZ, Leipzig, Germany}
\address[inst4]{Department of Biology, University of Maryland, College Park, MD, USA}

\begin{abstract}
Human mobility contributes to the fast spatio-temporal propagation of infectious diseases. During an outbreak, monitoring the infection situation on either side of an international border is very crucial as there is always a higher risk of disease importation associated with cross-border migration. Mechanistic models are effective tools to investigate the consequences of cross-border mobility on disease dynamics and help in designing effective control strategies. However, in practice, due to the unavailability of cross-border mobility data, it becomes difficult to propose reliable, model-based strategies. In this study, we propose a method for estimating cross-border mobility flux between any pair of regions that share an international border from the observed difference in the timing of the infection peak in each region. Assuming the underlying disease dynamics is governed by a Susceptible-Infected-Recovered (SIR) model, we employ stochastic simulations to obtain the maximum likelihood cross-border mobility estimate for any pair of regions where the difference in peak time can be measured. We then investigate how the estimate of cross-border mobility flux varies depending on the disease transmission rate, which is a key epidemiological parameter. We further show that the uncertainty in mobility flux estimates decreases for higher disease transmission rates and larger observed differences in peak timing. Finally, as a case study, we apply the method to some selected regions along the Poland-Germany border which are directly connected through multiple modes of transportation and quantify the cross-border fluxes from the COVID-19 cases data during the period $20^{\rm th}$ February $2021$ to $20^{\rm th}$ June $2021$.
\end{abstract}

\begin{keyword}
Coupling strength \sep Disease outbreak \sep Spatio-temporal model \sep Stochastic simulation \sep Maximum likelihood estimation 

\end{keyword}

\end{frontmatter}

\section{Introduction}
\label{sec:introduction}
In recent times, sporadic infectious disease outbreaks have created massive disruptions not only in public health functioning but also in socio-economic developments. Incomplete knowledge of disease transmission mechanisms, and unavailability of effective medical treatment cause difficulties in controlling disease spread. Advances in modern transportation technology further accelerate the spatial spread of infection and complicate the management of disease spread in affected regions~\citep{brockmann2009human, findlater2018human}. Mobility can occur on different spatial scales within a country. The complex interconnections among multiple scales limit the predictability of spatio-temporal patterns of disease spread~\citep{alessandretti2020scales, barbosa2018human,schlapfer2021universal}. Alternatively, inter-country migration often acts to introduce new pathogens in non-affected countries and within-country mobility can then quickly spread the disease throughout the country~\citep{wilder2008geographic}. Inter-country migration occurs primarily in two ways: international air-travel and daily commuting between countries sharing a border. The availability of digital technology makes it possible to keep track of the fluxes of global air-travel passengers. Moreover, during an outbreak, it helps to implement measures like airport screening, quarantining upon arrival, and contact tracing, which reduce the risk of disease import due to air travel. On the other hand, for daily cross-country mobility, due to a lack of convenient real-time database management~\citep{cavallaro2019cross}, it is often difficult to estimate the mobility flow, and, in turn, to quantify the component of infection risk because of human mobility.

Mathematical models are important tools for assessing the risk of disease import from a neighboring country and investigating the consequences of different control strategies. Reliable information on cross-border mobility flow not only hints at the effectiveness of intervention strategies like lockdowns or border-control measures but is also a key input parameter for multi-national epidemiological models in evaluating the performance of spatially-explicit intervention strategies. In very few cases, mobility flux information can be obtained from technology companies, public transit data, census data, and survey data. However, these data are very expensive to obtain, and often not publicly available due to privacy laws. Therefore, estimation of cross-country mobility flux remains a challenging task in the context of infectious disease modelling.

In spatial epidemiological analyses, mobility flux is usually accounted for via the gravity model, the radiation model, or related approaches~\citep{balcan2009multiscale,jandarov2014emulating, xia2004measles, senapati2019impact, mari2012modelling,yang2015transmission, mertel2022fine}. However, despite their wide use in the literature, these models have certain limitations as they rely on tuneable parameters which can vary with spatial location~\citep{simini2012universal}. Similarly, several theoretical frameworks have been proposed to reconstruct the underlying connection topology from the observed dynamics of the focal system~\citep{timme2007revealing, casadiego2017model,shandilya2011inferring}. These methods have been mainly applied to those systems that exhibit oscillatory dynamics, such as coupled chaotic oscillators~\citep{gao2022autonomous, ching2015reconstructing,shi2021inferring}, EEG time series data~\citep{alves2022eeg}, or neuronal systems~\citep{sysoeva2021reconstruction}. Recently, \cite{hempel2020estimating} proposed a method to estimate the coupling strength between two regions from invasion time, which is the time taken to reach the focal region after an infection is seeded in a neighboring region. In practice, however, the invasion time will not likely be a reliable enough basis for extracting mobility flow. This is because disease monitoring systems typically face the greatest difficulties in the early phase of a disease outbreak due to incompleteness or delays in reporting.  Consequently, all existing methods for estimating spatial connectivity are either data-hungry or depend heavily on precise information related to data reporting. A need therefore exists for a practical method that overcomes these limitations and can estimate cross-border mobility from imperfect monitoring data.

To this end, we propose an intuitive approach based on maximum-likelihood estimation to quantify the cross-border mobility flux for a pair of regions situated in two different countries. Our method assumes the underlying disease dynamics is governed by a simple Susceptible-Infected-Recovered (SIR) model and estimates the mobility flux from the observed difference in the timing of peak of infection between two regions. We first theoretically investigate, under realistic epidemiological assumptions, how the observed difference in peak-timing can effectively recover the underlying mobility between two regions and how the estimate of mobility varies depending on the disease transmission rate. Next, as a case study, we use our method to estimate the cross-country mobility flux from COVID-19 incidence data for the pair of regions which are connected directly through multiple modes of transportation and are located on either side of the Poland-Germany border.


\section{Methods}\label{sec:methods}
\subsection{Deterministic model}
We consider a standard deterministic two-patch SIR model incorporating short-term or commuting-type migration. Each spatial unit in our study, such as a district or even a state which shares border, can be considered a patch. In each patch $i$, the total population is divided into three classes based on health status: susceptible ($S_{i}$), infected ($I_{i}$), and recovered ($R_{i}$). Since the patches are connected through migration, the disease dynamics in each patch has two components: (i) disease dynamics within the patch, and (ii) disease dynamics between the patches. The mathematical equations describing the mechanism of disease dynamics in this two-patch scenario is given as follows:
\begin{eqnarray}
\begin{array}{lll}
\displaystyle \frac{d S_{i}}{dt} &=& -\displaystyle  \sum_{j=1}^{2}\beta_{j}m_{ij}S_{i}\Bigg(\frac{\displaystyle \sum_{k=1}^{2}m_{kj}I_{k}}{\displaystyle \sum_{k=1}^{2}m_{kj}N_{k} } \Bigg),\\

\displaystyle \frac{d I_{i}}{dt} &=& \displaystyle  \displaystyle  \sum_{j=1}^{2}\beta_{j}m_{ij}S_{i}\Bigg(\frac{\displaystyle \sum_{k=1}^{2}m_{kj}I_{k}}{\displaystyle \sum_{k=1}^{2}m_{kj}N_{k} } \Bigg) - \gamma_{i} I_{i},\\

\displaystyle \frac{d R_{i}}{dt} &=& \displaystyle  \gamma_{i} I_{i}, \hspace*{1cm}~i~=~1,~2.
\end{array}
\label{EQ:Model}
\end{eqnarray}
The rate at which the susceptible population in patch $i$ gets infected in patch $j$ is given by the product of three terms:  the risk of infection in patch $j$ ($\beta_{j}$), the number of susceptibles from patch $i$ who are currently in patch $j$ ($m_{ij}S_{i}$), and the proportion of the population that is infected in patch $j$ ($\frac{\sum_{k=1}^{2}m_{kj}I_{k}}{ \sum_{k=1}^{2}m_{kj}N_{k} }$). Here, $N_{j}$ denotes the total population in patch $j$ and $\sum_{k=1}^{2}m_{kj}N_{k}$ gives the total population currently present in patch $j$ is called effective population of patch $j$~\citep{bichara2015sis}. The model,~\eqref{EQ:Model}, has two epidemiological parameters: transmission rate ($\beta_{i}$), and infectious period ($\frac{1}{\gamma_{i}}$). The matrix $m_{ij}$ is called the residence-time matrix, where the element $m_{ij}$ can be interpreted as the fraction of people from patch $i$ that visited patch $j$~\citep{bichara2015sis}. Since $m_{ij}$ denotes the fraction of mobility flux, we have the constraint $\sum_{j=1}^{2}m_{ij}=1$, for $i=1,2$. 

For a fixed set of model parameters and initial conditions, we define $t_{d}$ as the difference between the peak timing of these two regions, which can be expressed: 
\begin{equation*}
    t_{d}= |t_{1}^{\max} - t_{2}^{\max}|,
\end{equation*}
    where, $t_{1}^{\max}$ and $t_{2}^{\max}$ denote the time points at which the trajectories of the infected populations $I_{1}(t)$ and $I_{2}(t)$ reach their respective maxima.  
    
\subsection{Stochastic model}
\label{subsec:stochastic_model}
Here we are interested in estimating the cross-border mobility flux from the observed difference in peak timing, $t_d$. It should be noted that in a deterministic setup, for given model parameters and initial conditions, we will always end up with exactly one $t_{d}$. Therefore, we introduce randomness into the model as a form of demographic stochasticity. Here demographic stochasticity refers to the fluctuations in the disease propagation process arising from the random nature of disease transmission and recovery events at the individual level. We consider a standard event-driven approach called the \textit{tau-leaping} method~\citep{gillespie2001approximate}. This method is a modified version of the Gillespie algorithm where the inter-event duration is kept fixed instead of following an exponential distribution, which increases computational efficiency~\citep{gillespie2001approximate,gillespie2003improved}.

Let us denote the number of transmission and recovery events at time $t$ for each patch $i$ (i=1, 2) by $\mathcal{E}_{\rm Trans}^{i}(t)$ and $\mathcal{E}_{\rm Rec}^{i}(t)$ respectively. The transition probabilities for the events $\mathcal{E}_{\rm Trans}^{i}(t)$ and $\mathcal{E}_{\rm Rec}^{i}(t)$ that occur during the small but fixed time interval $\delta t$ can be written as

\begin{eqnarray}
\begin{array}{lll}
\mathbb{P}(\delta \mathcal{E}_{\rm Trans}^{i}=1|S_{i}, I_{i}, R_{i})=\displaystyle  \displaystyle  \sum_{j=1}^{2}\beta_{j}m_{ij}S_{i}\Bigg(\frac{\displaystyle \sum_{k=1}^{2}m_{kj}I_{k}}{\displaystyle \sum_{k=1}^{2}m_{kj}N_{k} } \Bigg) \delta t + o(\delta t),\\

\mathbb{P}(\delta \mathcal{E}_{\rm Rec}^{i}=1|I_{i})= \gamma_{i} I_{i} \delta t + o(\delta t),
\end{array}
\end{eqnarray}
where, $\delta \mathcal{E}_{\rm Trans}^{i}=\mathcal{E}_{\rm Trans}^{i}(t+\delta t) - \mathcal{E}_{\rm Trans}^{i}(t),$ and 
 $\delta \mathcal{E}_{\rm Rec}^{i}=\mathcal{E}_{\rm Rec}^{i}(t+\delta t) - \mathcal{E}_{\rm Rec}^{i}(t)$ for $i=1,2$. 
 
For small $\delta t$, the increments $\delta \mathcal{E}_{\rm Trans}^{i}$ and $\delta \mathcal{E}_{\rm Rec}^{i}$ are approximated as Poisson distributed random variables: 
\begin{eqnarray}
\begin{array}{lll}
\delta \mathcal{E}_{\rm Trans}^{i}\approx Poisson(\displaystyle  \sum_{j=1}^{2}\beta_{j}m_{ij}S_{i}\Bigg(\frac{\displaystyle \sum_{k=1}^{2}m_{kj}I_{k}}{\displaystyle \sum_{k=1}^{2}m_{kj}N_{k} } \Bigg) \delta t ),\\

\delta \mathcal{E}_{\rm Rec}^{i}\approx Poisson(\gamma_{i}I_{i} \delta t). 
\end{array}
\end{eqnarray}

The population sizes in different compartments are then updated as, 
\begin{eqnarray}
\begin{array}{lll}
S_{i}(t+\delta t)= S_{i}(t) -\delta \mathcal{E}_{\rm Trans}^{i} ,\\
I_{i}(t+\delta t)= I_{i}(t) +\delta \mathcal{E}_{\rm Trans}^{i} - \delta \mathcal{E}_{\rm Rec}^{i},\\
R_{i}(t+\delta t)=R_{i}(t) +\delta \mathcal{E}_{\rm Rec}^{i}.
\end{array}
\end{eqnarray}

\begin{figure}[H]
    	\begin{center}
    	\includegraphics[width=1\textwidth]{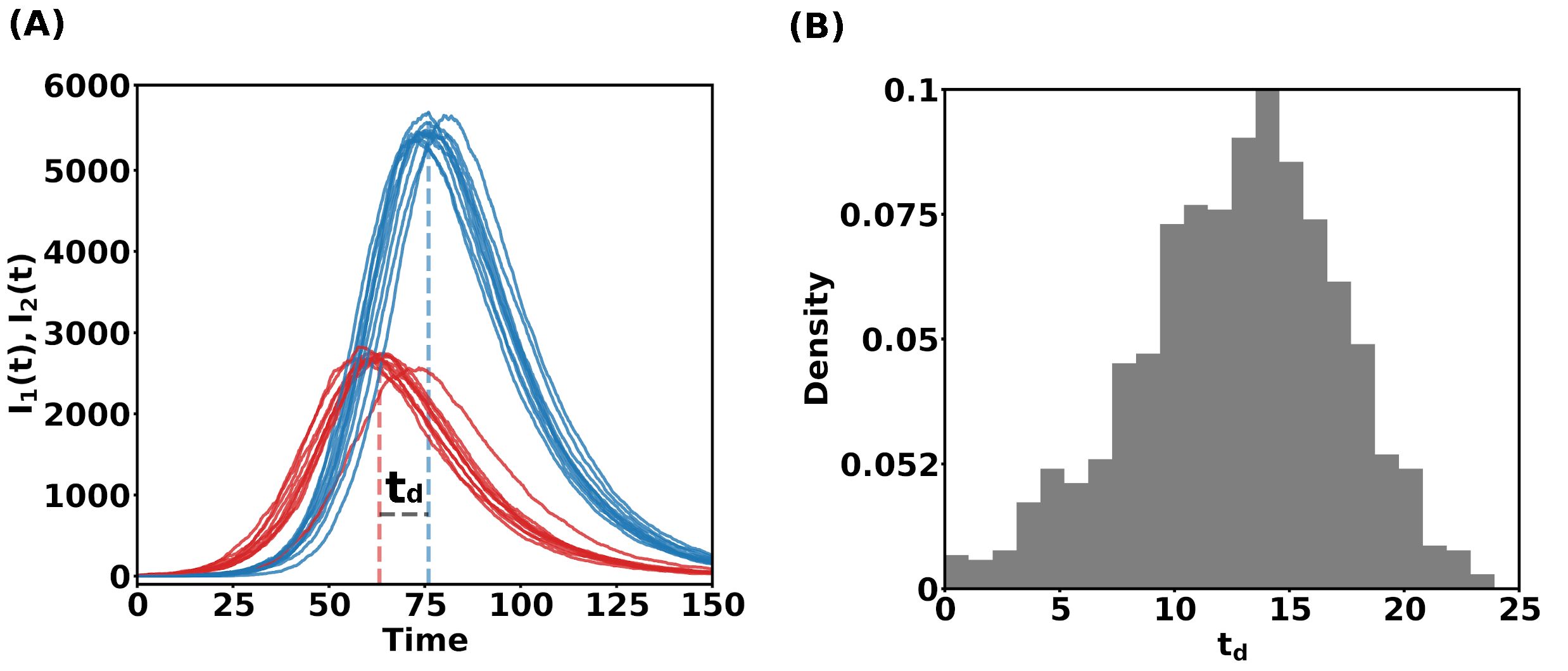}
    	\end{center}
    	\caption{ (A) Example of stochastic realizations of the model for a fixed combination of model parameters. The red and blue trajectories refer to $I_{1}(t)$, and $I_{2}(t)$ respectively. The dashed red and blue line indicate the points at which $I_{1}(t)$ and $I_{2}(t)$ attain their maxima respectively. The difference between these time points is shown as $t_{d}$. (B) Histogram of $t_{d}$ for 10,000 stochastic realizations.    }
    	\label{fig:fig_schematic}
\end{figure}

Given model parameters $\beta_{i}, \gamma_{i}, N_{i}, I_{i}(0)$, for a particular residence-time matrix $m_{ij}$, we generate multiple stochastic realizations. In each realization, we calculate the difference in peak timing ($t_{d}$) and consequently the distribution of it. We show an example in Fig.~\ref{fig:fig_schematic} (A)-(B), where we consider the parameters as $\beta_1=0.30$, $\beta_2=0.20$, $\gamma_1=\gamma_2 = 0.07$, $N_1=10000$, $N_2=20000$, $I_{1}(0)=10$, $I_{2}(0)=1$ and generate multiple stochastic realizations. In each of the realizations, we calculate $t_d$ (see Fig.~\ref{fig:fig_schematic} (A)) and plot its distribution as a histogram ( see Fig.~\ref{fig:fig_schematic} (B)). 

To produce the likelihoods, we first consider a range of values for the mobility parameter and divide it into 100 linearly-spaced values. Then for each value of mobility parameter, we produce 10000 stochastic simulations and for each simulation, we calculate $t_d$. The full range of $t_d$ values is then divided into 100 bins. Then we calculate the number of simulations for which the corresponding bin contains the observed $t_d$ and consequently, this gives the likelihood of mobility parameter given $t_d$. Finally, we take the logarithm of these likelihoods (i.e. log-likelihoods (LL)) and calculate the maximum likelihood estimate (MLE) of mobility parameter given $t_d$.

\section{Results}\label{sec:results}
We first investigate the relationship between difference in peak timing ($t_{d}$) and cross-country mobility ($m_{ij}$). For the sake of simplicity we assume the mobility in both  directions is the same, i.e, $m_{12}$=$m_{21}$=$m$. We also take the recovery rates to be the same for both the regions, $\gamma_{1}=\gamma_{2}=\gamma$. We fix $\gamma=\frac{1}{14}~day^{-1}$ throughout our study unless specified otherwise. For the transmission rate, we consider two cases: (i) $\beta_{1}$ and $\beta_{2}$ are equal, i.e $\beta_1 =\beta_2=\beta$, and (ii) $\beta_1$ and $\beta_2$ are not equal. We consider the total population $N_{1}=10000$ and $N_{2}=20000$, which is fixed over time, and we assume initially there is no recovered individual in either region, i.e $R_{1}(0)=0$, and $R_{2}(0)=0$. 

\begin{figure}[H]
    	\begin{center}
    	\includegraphics[width=1\textwidth]{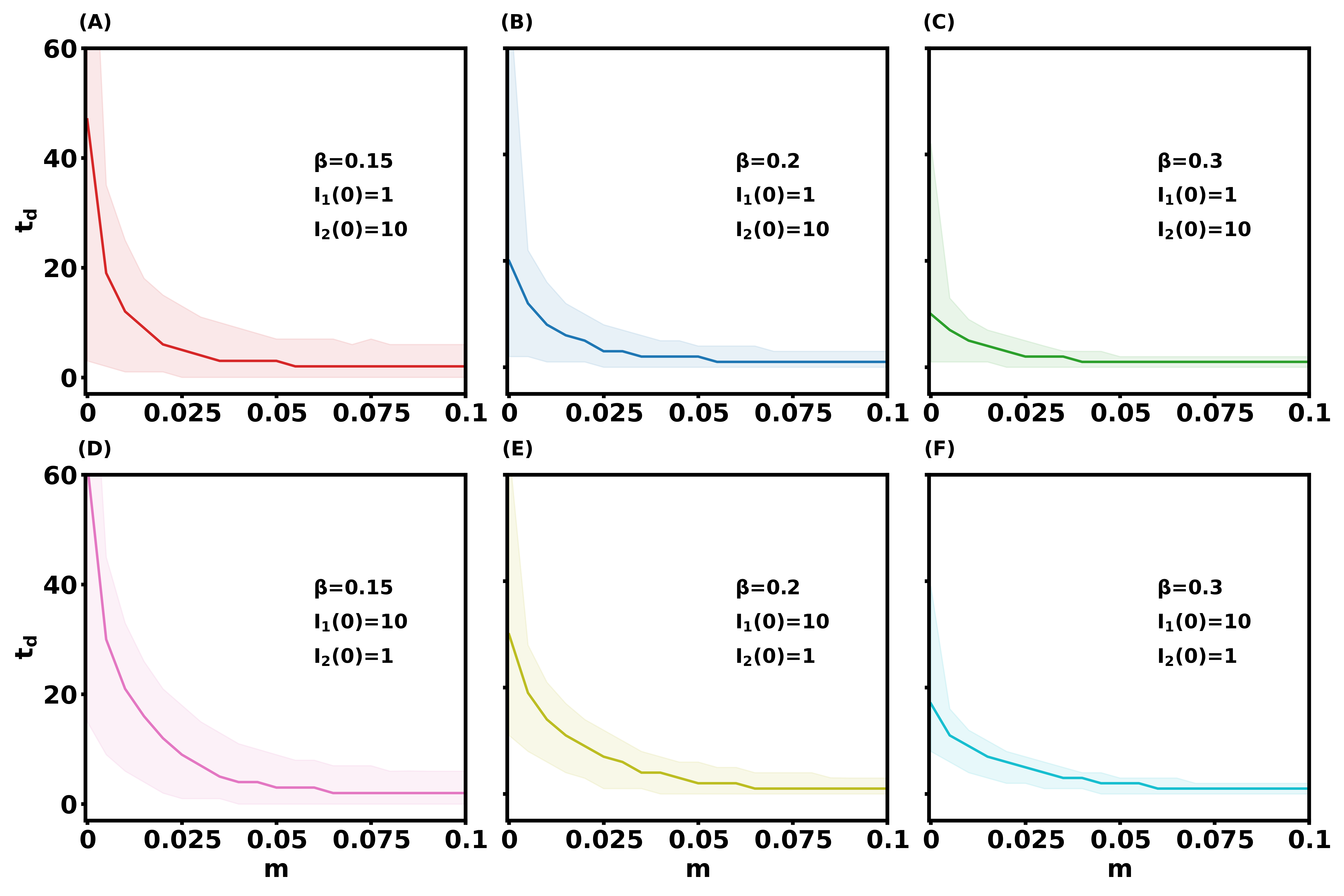}
    	\end{center}
    	\caption{(A)-(F)Behaviour of difference in peak-timing ($t_{d}$) with respect to coupling strength ($m$) for different combinations of transmission rate and initial infection. Here, the solid line represents the mean of the stochastic realizations and the shaded region represents $95\%$ confidence interval around the mean.}
    	\label{fig:fig_m_vs_td}
\end{figure}

Following the method described in Section~\ref{subsec:stochastic_model}, for each value of the mobility parameter ($m$), we take $10,000$ realizations and calculate $t_d$ for different combinations of initial infected population ($I_{1}(0)$, and $I_{2}(0)$) and disease transmission rate ($\beta$). We see that the difference in peak timing ($t_{d}$) gradually decreases with increasing mobility flux $m$ (see Fig.~\ref{fig:fig_m_vs_td}(A)-(F)). 

\begin{figure}[H]
    	\begin{center}
    	\includegraphics[width=1\textwidth]{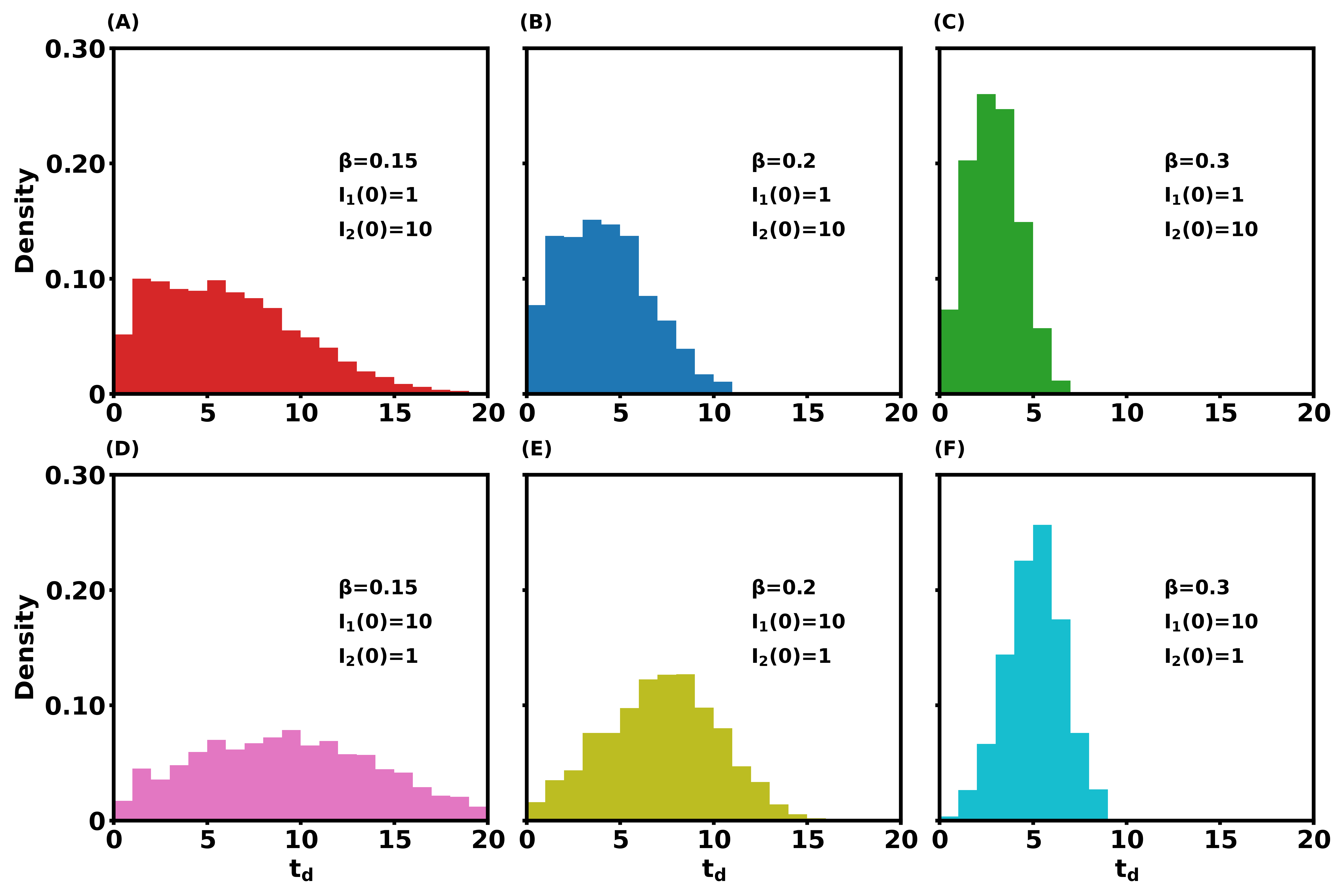}
    	\end{center}
    	\caption{(A)-(F) Distribution of $t_d$ corresponding to the mobility parameter $m=0.025$, for different parameter combinations as mentioned in Fig~\ref{fig:fig_m_vs_td}.   }
    	\label{fig:fig_distri_td}
\end{figure}

We see that for a particular value of $m$, the difference in peak timing is higher for low disease transmission rate, and if we increase the transmission rate it starts to decrease. For instance, if we fix initial infecteds as $I_{1}(0)=1$, and $I_{2}(0)=10$, we see that for lower transmission rate $\beta=0.15$, and mobility parameter $m=0.025$, the difference in peak timing varies in the range [0,18] days (see Fig.~\ref{fig:fig_distri_td}(A)). If we slightly increase the transmission rate, i.e $\beta=0.2$, we observe that the range of $t_{d}$ shrinks to [0,10] days (see Fig.~\ref{fig:fig_distri_td}(B)). For higher transmission rate ($\beta=0.3$), the range of $t_d$, in this case, reduces to [0,6] days (see Fig.~\ref{fig:fig_distri_td})(C). Similar patterns are observed even if we change the number of initially infected individuals. For $I_{1}(0)=10$, and $I_{2}(0)=1$, we see, for $\beta=0.15$, the range of $t_{d}$ is [0,20] (see Fig.~\ref{fig:fig_distri_td}(D)). It shrinks to [0,14] for $\beta=0.2$ and finally to [1,9] for $\beta=0.2$ (see Fig.~\ref{fig:fig_distri_td}(E)-(F)). We also note that for low transmission rate values, the distribution of $t_d$ is flatter than that for higher transmission rates (see Fig.~\ref{fig:fig_distri_td}). This indicates that estimation of the mobility parameter from observed differences in peak timing ($t_d$) would be associated with greater uncertainty in the scenario where the transmissibility of the infection is comparatively lower.  

Now we explore the feasibility of estimating the mobility flux ($m$) from observed differences in peak timing ($t_d$). We calculate the maximum likelihood estimate (MLE) of the parameter $m$, given observed value of $t_d$, and other model parameters ($\beta, \gamma,N_{1},N_{2}, I_{1}(0), I_{2}(0)$). Here, observed value of $t_d$ in each case is obtained by integrating the deterministic model~\eqref{EQ:Model} with same set of fixed parameters. Since it is evident from the above analysis that the transmission rate ($\beta$) is influential in calculating $t_{d}$, for an observed value $t_d$, we consider two different values of $\beta$ and estimate the MLE of the parameter $m$. 

\begin{figure}[H]
    	\begin{center}
    	\includegraphics[width=1\textwidth]{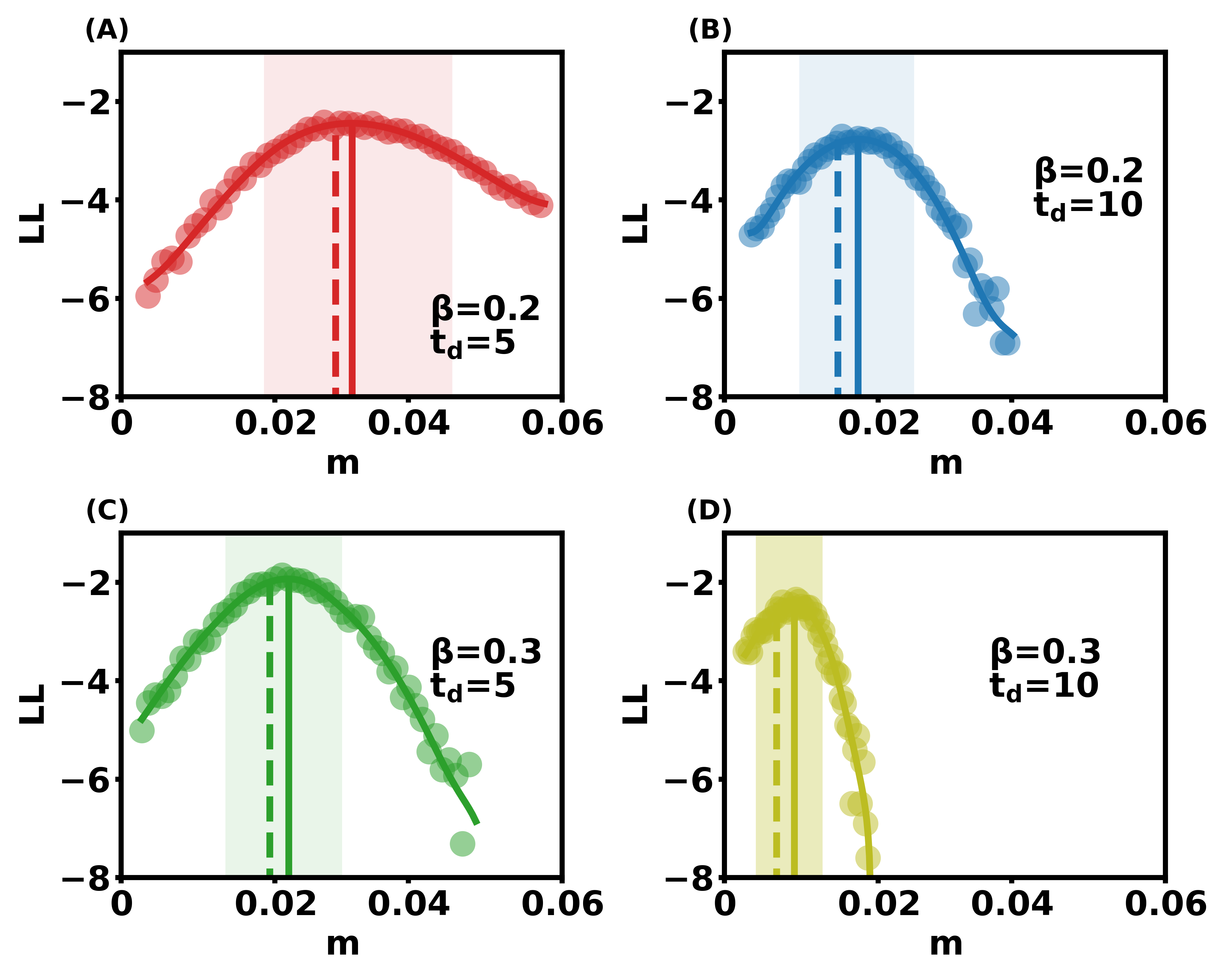}
    	\end{center}
    	\caption{ (A)-(D) Maximum likelihood estimates of the mobility parameter ($m$) given the difference in peak timing ($t_d$). The solid circles are the log-likelihood obtained from stochastic simulations and the solid curve is the smoothed likelihood profile. The solid vertical line is the maximum likelihood estimate of $m$ and the dashed vertical line is the true value of $m$, for which the deterministic model~\eqref{EQ:Model} gives the corresponding observed value of $t_{d}$. The shaded region depicts the $75\%$ confidence interval. The remaining parameters are fixed as $\gamma=\frac{1}{14}$, $N_{1}=10000$, $N_{2}=20000$, $I_{1}(0)=10$, $I_{2}(0)=1$.   }
    	\label{fig:mle_theory}
\end{figure}

We calculate the log-likelihood (LL) of the parameter $m$ for a range of possible values using the stochastic simulation described in Section~\ref{subsec:stochastic_model}. Then using cubic-splines, we smooth the likelihood profile to obtain the maximum likelihood estimate and the corresponding confidence interval using a likelihood ratio test. 
For example, we consider a case where the observed value of $t_{d}$ is $5$ days. If we fix $\beta=0.2$, then MLE of $m$ is $0.031$ and if $\beta$ is taken as $0.3$, the MLE of $m$ is reduced to $0.021$ (see Fig.~\ref{fig:mle_theory}(A), (C)). In another example, we consider higher observed value of $t_{d}$, i.e 10 days. In this case, for $\beta=0.2$ the MLE of $m$ is $0.018$ and for $\beta=0.3$ it becomes $0.01$ (see Fig.~\ref{fig:mle_theory}(B),(D)). From Fig.~\ref{fig:mle_theory}(A)-(D), in all four cases, we can see that MLEs of $m$ (vertical solid lines) are quite close to the true value of $m$ (dashed vertical line), the value of mobility parameter obtained from~\eqref{EQ:Model} for the corresponding value of $t_{d}$. 

Note also that for a given $t_{d}$, the uncertainty in estimating the mobility parameter decreases with increasing disease transmission rate $\beta$. For both the lower and higher observed $t_{d}$ scenarios, the confidence intervals on $m$ shrink when we increase $\beta$ (see Fig.~\ref{fig:mle_theory}(A), (C) and (B), (D)). For a fixed value of $\beta$, if the observed value of $t_d$ is higher, then we also observe reduced uncertainty. For example, if $\beta$ is fixed as $0.2$, then the confidence interval on $m$ is wider for $t_d=5$ days than $t_d=10$ (see Fig.~\ref{fig:mle_theory}(A), (B)). A similar trend is also observed for higher transmission rates, e.g. $\beta=0.3$ (see Fig.~\ref{fig:mle_theory} (C), (D)). 

\begin{figure}[H]
    	\begin{center}
    	\includegraphics[width=1\textwidth]{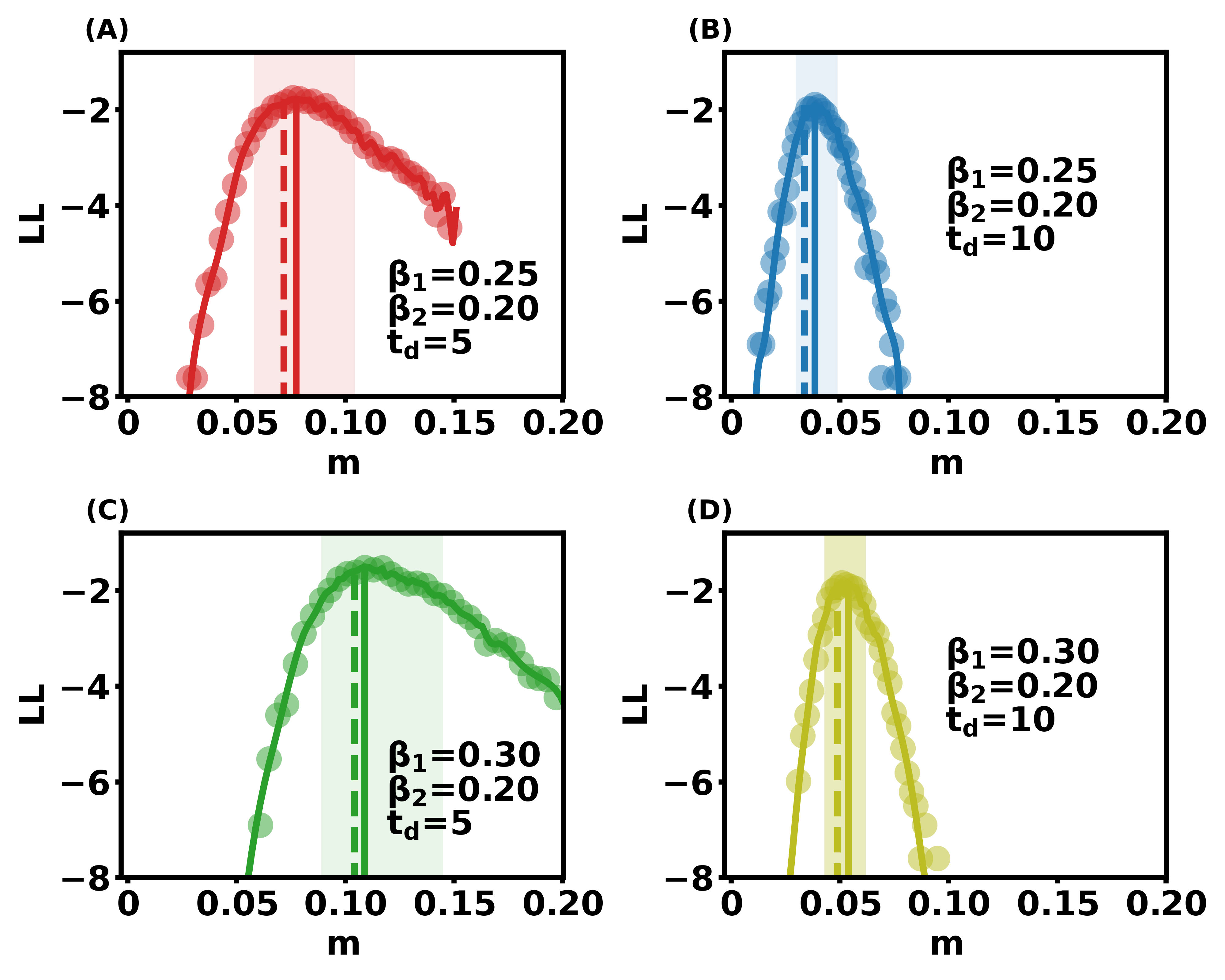}
    	\end{center}
    	\caption{ (A)-(D) Maximum likelihood estimate of the mobility parameter ($m$) given the difference in peak timing ($t_d$), when $\beta_{1}\neq \beta_2$. The solid circles are the log-likelihood obtained from stochastic simulation and the solid curve is the smoothed likelihood profile. Solid vertical line is the maximum likelihood estimate of $m$ and the dashed vertical line is the true value of $m$, for which the deterministic model~\eqref{EQ:Model} gives the corresponding observed value of $t_{d}$. The shaded region depicts the $75\%$ confidence interval. The remaining parameters are fixed as $\gamma=\frac{1}{14}$, $N_{1}=10000$, $N_{2}=20000$, $I_{1}(0)=10$, $I_{2}(0)=1$.   }
    	\label{fig:mle_theory_diff_beta}
\end{figure}

Now we consider the case when $\beta_{1} \neq \beta_{2}$. We will explore how the estimate of the mobility parameter changes with the difference in transmission rates. To illustrate this, we consider a case where the observed value of $t_{d}$ is $5$ days. If we fix $\beta_1=0.25$, and $\beta_2 =0.20$ then MLE of $m$ is $0.076$ and if $\beta_1$, and $\beta_2$ are taken as $0.30$, and $0.20$ respectively, the MLE of $m$ is increased to $0.108$ (see Fig.~\ref{fig:mle_theory_diff_beta}(A), (C)). In another example, we consider a larger observed value of $t_{d}$, i.e 10 days. In this case, for $\beta_1=0.25$, and $\beta_2=0.20$, the MLE of $m$ is $0.038$ and for $\beta_1=0.30$, and $\beta_2 = 0.20$, it becomes $0.051$ (see Fig.~\ref{fig:mle_theory_diff_beta}(B),(D)). Similar to the case when $\beta_1=\beta_2$, from Fig.~\ref{fig:mle_theory_diff_beta}(A)-(D), we can see that MLEs of $m$ (vertical solid lines) are quite close to the true value of $m$ (dashed vertical line), the value of mobility parameter obtained from~\eqref{EQ:Model} for the corresponding value of $t_{d}$. 

\subsection*{Case study: COVID-19 incidence in Poland-Germany border region}
\begin{figure}[H]
    	\begin{center}
    	\includegraphics[width=1\textwidth]{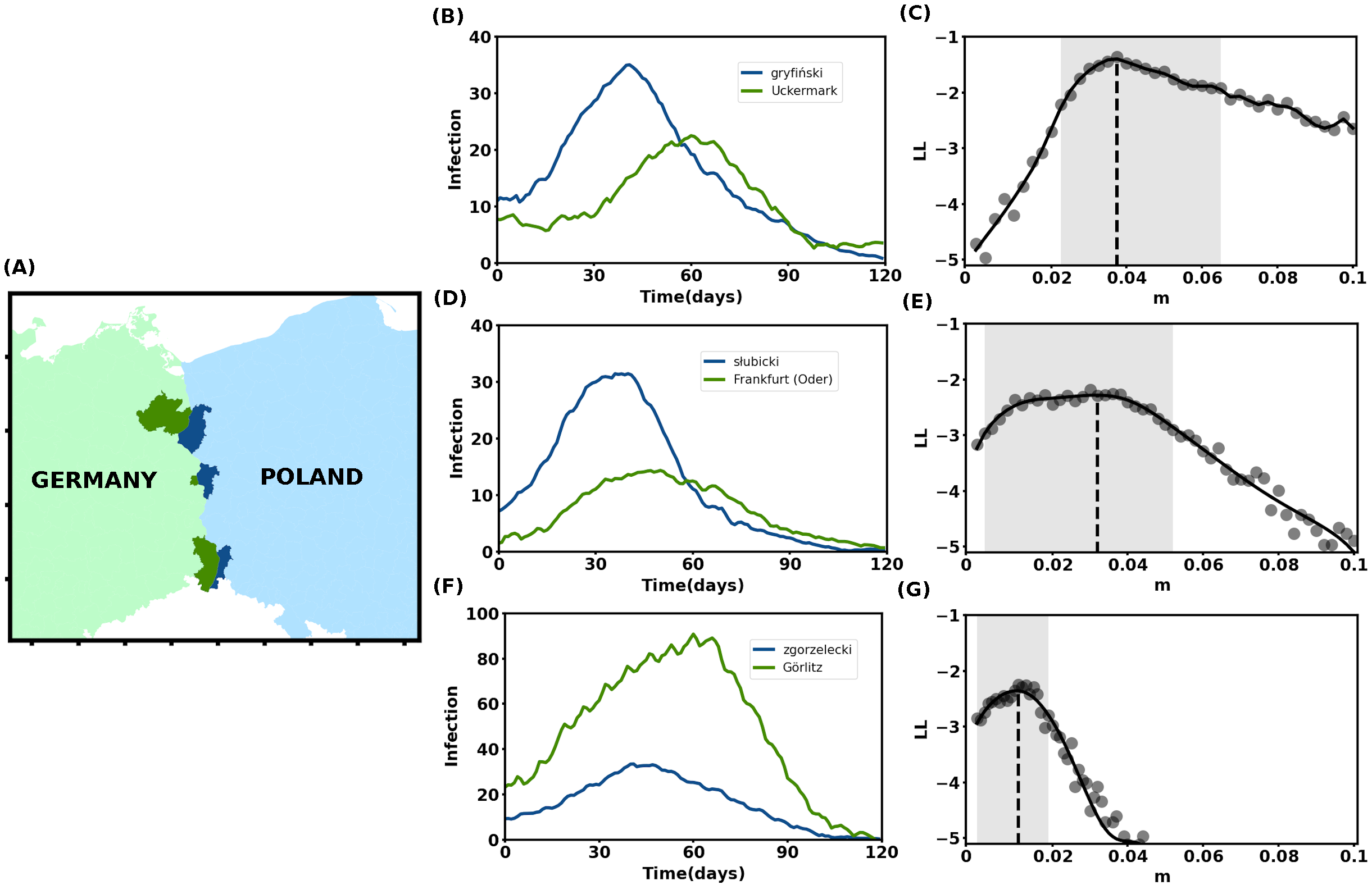}
    	\end{center}
    	\caption{(A) The map shows the three pairs of regions that share a border between Poland and Germany. The Polish regions (called \textit{Powiat}) are filled with dark blue color, and the German regions (called \textit{Kreise}) are shaded dark green. The name of the regions are: top-(Gryfi\'{n}ski, Uckermark); middle- (S\l{}ubicki, Frankfurt(Oder)); and bottom-(Zgorzelecki,G\"{o}rlitz). (B), (D), (F) The time series of number of infecteds in the selected regions. The blue lines represent the time series for the Polish regions and green lines represent the time series for the German regions. Here, time zero refers to the date $20^{\rm th}$ February 2021 and time 120 refers to the date $20^{\rm th}$ June 2021. (C), (E), (G) The likelihood profiles of the cross-border mobility $m$. The gray circles are the log-likelihoods obtained from the stochastic simulation and the solid curve is the smoothed likelihood profile. The dashed vertical line is the maximum likelihood estimate of $m$ and the shaded region denotes the $75\%$ confidence interval.  }
    	\label{fig:mle_data}
\end{figure}

We apply our proposed method to COVID-19 incidence data during the period $20^{\rm th}$ February 2021 to $20^{\rm th}$ June 2021, for the pair of regions which are located on either side of the Poland-Germany border and are directly connected through different modes of transportation e.g., bus-transportation, train-transportation (see~Fig.\ref{fig:mle_data} (A)). For each pair of regions, we obtain the corresponding information on total population, and initial number of infections (see Table~\ref{table:table_case_study}). All these datasets are obtained from the database pipeline by~\cite{abdussalam2022scalable}. We fix the value of average infectious period ($\frac{1}{\gamma}$) at $14$, following ~\cite{byrne2020inferred}.

We have previously demonstrated that the disease transmission rate plays an influential role in the estimation of mobility flux given the difference in peak-timing. Therefore, in the case of real data, instead of fixing the disease transmission rate arbitrarily, we follow a simpler approach to estimate these value from the initial phase of the epidemic. Since, early in the epidemic the daily number of new cases grows exponentially over time, for a SIR-type model the number of infection at time $t$ can be approximated as~\citep{ma2020estimating}: 
\begin{equation*}
    I(t)\approx I(0)\exp{(\lambda t)}.
\end{equation*}
Here, $\lambda$ is the growth rate and is given by $\lambda=\beta - \gamma$. Since cumulative number of new cases per day ($C(t)$) and new cases per day ($I(t)$) are linearly related~\citep{favier2006early, pinho2010modelling}, i.e $I(t)$ $\sim$ $\lambda$ $C(t)$, we estimate the growth rate $\lambda$ using a linear fit. Therefore, once the parameter $\gamma$ is kept fixed, the disease transmission rate $\beta$ can be approximated as $\beta \approx \lambda + \gamma$. Following this approach, for each region we estimate the disease transmission rate from the time series data of daily new cases (see Table~\ref{table:table_case_study}). 

 \begin{table}[H]
  \tiny
  	\centering
  	\begin{tabular}{|p{3.1cm}|p{1.5cm}|p{1.5cm}|p{1.3cm}|p{1.3cm}|p{1.9cm}|}
  		\hline
  		\centering{\textbf{Pair (Powiat, Kreise)}} &\centering{\textbf{Transmission rate ($\beta_{1}$, $\beta_2$)} } &\centering{\textbf{Total Population ($N_{1}$, $N_{2}$)}} & \textbf{Initial infection $(I_{1}(0), I_{2}(0))$} & \textbf{Observed difference in peak-timing ($t_{d}$)} & \textbf{Estimated mobility flux ($m[75\% CI]$)} \\
  		[2ex]
  		\hline
  		\centering (Gryfi\'{n}ski, Uckermark) & (0.10,0.075) & (82951,119552)  & (11,8)  & 20 days & 0.038[0.022,0.064]\\[5ex]
  		\centering (S\l{}ubicki, Frankfurt(Oder)) & (0.13,0.14)  & (47068,57873) & (8,1) & 10 days & 0.032[0.004,0.052]\\[5ex]
  		\centering(Zgorzelecki,G\"{o}rlitz) & (0.10,0.11) & (90584,254894) & (9,23) & 20 days  & 0.012[0.002,0.018]\\[4ex]
  		
  		\hline
  	\end{tabular}
  	\caption{Parameters and observed difference in peak-timing ($t_{d}$) for each pair of region used for estimating the mobility flux $m$. The estimated value of $m$ is presented in the last column of the table. }
  	\label{table:table_case_study}
  \end{table}

We first consider the regions: Gryfi\'{n}ski, and Uckermark, which are located at the upper part of the Poland-Germany border. The observed difference in peak-timing obtained from the real data is $20$ days (see Fig.~\ref{fig:mle_data}(B)). Given all the other model parameters, we estimate the mobility flux as $0.038 [0.022,0.064]$ (see Fig.~\ref{fig:mle_data}(C)). Next, we move to the middle part of the border and choose the regions: S\l{}ubicki, and Frankfurt(Oder). In this case, the observed difference in peak-timing is $10$ days (see Fig.~\ref{fig:mle_data} (D)). The maximum likelihood estimate of $m$ is $0.032[0.004,0.052]$ (see Fig.~\ref{fig:mle_data} (E)). Now for the regions: Zgorzelecki, and G\"{o}rlitz, the observed difference in peak timing is $20$ days (see Fig.~\ref{fig:mle_data} (F)). The estimated value of the parameter $m$ in this case is $0.012[0.002,0.018]$ (see Fig.~\ref{fig:mle_data} (G)). 

Comparing the first case and third case, we see that there is a substantial difference between the estimates of mobility parameters even though the observed differences in peak-time ($t_{d}$) for both cases are same. Instead, the difference in estimated transmission rates (i.e, $\beta_{1}$ and $\beta_2$) is higher in the first case (Gryfi\'{n}ski, and Uckermark) than that of the third case (Zgorzelecki, and G\"{o}rlitz) (see Table~\ref{table:table_case_study}). This marked difference in transmission rates likely explains the different estimates of the coupling parameter. 

\section{Discussion}\label{sec:discussion}
In general, understanding the interplay between human mobility and spatio-temporal propagation of infectious diseases is a long-standing problem in the field of infectious disease modelling. The unavailability of reliable information on human mobility poses challenges to designing effective public health strategies. In this study, we proposed a method to address the issue of estimating cross-country mobility flux during a disease outbreak, which is important yet less explored in the literature. We showed that from the observed  difference in peak-timing between any two regions which share a border, the underlying mobility flow between them can be retrieved effectively. Since, the information on peak timing is less affected than the invasion time by the delay or incompleteness in reporting infection, therefore the difference in peak-timing can be considered as a more robust quantity than invasion time. Consequently, our approach would be more broadly applicable for estimating mobility flux in real-world case studies than existing alternatives that rely on the observed disease invasion time~\citep{hempel2020estimating}.

 We investigated how different model parameters influence the difference in peak-timing in a simple scenario, where the disease transmission rates in both regions are the same. We see that the disease transmission rate plays key role in determining the distribution of difference in peak-timing. The difference in peak-timing decreases with the disease transmission rate (Fig.~\ref{fig:fig_m_vs_td}). This is because of the fact that higher disease transmission rate causes earlier occurence of the peak in infection trajectory and thus the difference in peak-timing between the two regions also decreases. It has also been revealed that, for lower transmission rates, the distribution of differences in peak-timing becomes flatter and can have a wider range of possible values (Fig.~\ref{fig:fig_distri_td}(A), (D)). On the other hand, with increasing transmission rates the distribution progressively narrows (Fig.~\ref{fig:fig_distri_td} (B), (C), (E), (F)). 

Based on a simple SIR-type model, with the help of stochastic simulations, we provided maximum likelihood estimate of the cross-border mobility under different epidemiologically relevant scenarios. From our numerical investigation, it is evident that the estimation method based on the observed difference in peak-timing can effectively recover the underlying mobility rate for both scenarios: (i) when the disease transmission rates in both regions are assumed to be the same (Fig.~\ref{fig:mle_theory}(A)-(D)), and (ii) when the transmission rates are assumed to be different (Fig.~\ref{fig:mle_theory_diff_beta}(A)-(D)). In the first scenario, for a given difference in peak-timing, the uncertainty of the estimate decreases with the transmission rate (Fig.~\ref{fig:mle_theory}(A), (C) and Fig.~\ref{fig:mle_theory}(B), (D)). Also for a fixed transmission rate, the uncertainty in estimation decreases with increasing values of the observed difference in peak timing (Fig.~\ref{fig:mle_theory}(A), (B) and Fig.~\ref{fig:mle_theory}(C), (D)). In the second scenario, for a given observed difference in peak-timing, the estimate of mobility parameter increases with the difference between the disease transmission rates (Fig.~\ref{fig:mle_theory_diff_beta}(A), (C) and Fig.~\ref{fig:mle_theory_diff_beta}(B), (D)). Similar to the first scenario, for fixed disease transmission rates, the uncertainty in the estimate decreases with the observed difference in peak-timing (Fig.~\ref{fig:mle_theory_diff_beta}(A), (B) and Fig.~\ref{fig:mle_theory_diff_beta}(C), (D)). 

To demonstrate its real-world utility, we applied our proposed method to COVID-19 incidence data obtained for the regions located along the border of Poland and Germany. We chose only three pairs of regions located in different parts of the Polish-German border which are well connected through different modes of transportation. However, the method can also be applied to any other pair of regions that shares border. Using least squares techniques, we first estimated the disease transmission rate for each of the regions from the early phase of epidemics. We then calculated the maximum likelihood estimate of the mobility parameter for three pairs of regions located in different areas along the border. It is noteworthy that, if we compare the pair of regions Gryfi\'{n}ski, Uckermark and Zgorzelecki, G\"{o}rlitz, we see that, even though the differences in peak-timing are the same, the estimated values for the cross-border mobility are very different. Since the difference in transmission rates between the regions Gryfi\'{n}ski and Uckermark is higher than that of Zgorzelecki and G\"{o}rlitz, we see higher mobility flow for the regions Gryfi\'{n}ski and Uckermark. This result is in good agreement with our theoretical findings (Fig.~\ref{fig:mle_theory_diff_beta}).

Even though our study addressed the important issue of estimation of cross-border mobility and our proposed method worked well in the real case study, it has some limitations. We assumed that the mobility between two regions is symmetric, which is not true in general, specifically if we consider the mobility between a big city and a village. We considered the regions on either side of the border as isolated and ignored the contribution of the adjacent regions which is not realistic. We leave these modifications and extensions for future study.

Summing up, this work provides a simple and intuitive method for estimating mobility flow between the regions connected through human migration. The methodology and findings can serve as a basis for designing intervention strategies in situations where direct information on human mobility is inaccessible.

\section*{Acknowledgements}
This work was partially funded by the Center of Advanced Systems Understanding (CASUS), which is financed by Germany’s Federal Ministry of Education and Research (BMBF) and by the Saxon Ministry for Science, Culture, and Tourism (SMWK) with tax funds on the basis of the budget approved by the Saxon State Parliament.

\bibliographystyle{elsarticle-harv} 
\bibliography{main}

\begin{thebibliography}{32}
\expandafter\ifx\csname natexlab\endcsname\relax\def\natexlab#1{#1}\fi
\providecommand{\url}[1]{\texttt{#1}}
\providecommand{\href}[2]{#2}
\providecommand{\path}[1]{#1}
\providecommand{\DOIprefix}{doi:}
\providecommand{\ArXivprefix}{arXiv:}
\providecommand{\URLprefix}{URL: }
\providecommand{\Pubmedprefix}{pmid:}
\providecommand{\doi}[1]{\href{http://dx.doi.org/#1}{\path{#1}}}
\providecommand{\Pubmed}[1]{\href{pmid:#1}{\path{#1}}}
\providecommand{\bibinfo}[2]{#2}
\ifx\xfnm\relax \def\xfnm[#1]{\unskip,\space#1}\fi
\bibitem[{Abdussalam et~al.(2022)Abdussalam, Mertel, Fan, Sch{\"u}ler,
  Schlechte-We{\l}nicz and Calabrese}]{abdussalam2022scalable}
\bibinfo{author}{Abdussalam, W.}, \bibinfo{author}{Mertel, A.},
  \bibinfo{author}{Fan, K.}, \bibinfo{author}{Sch{\"u}ler, L.},
  \bibinfo{author}{Schlechte-We{\l}nicz, W.}, \bibinfo{author}{Calabrese,
  J.M.}, \bibinfo{year}{2022}.
\newblock \bibinfo{title}{A scalable pipeline for {COVID-19}: the case study of
  {G}ermany, {C}zechia and {P}oland}.
\newblock \bibinfo{journal}{arXiv preprint arXiv:2208.12928} .
\bibitem[{Alessandretti et~al.(2020)Alessandretti, Aslak and
  Lehmann}]{alessandretti2020scales}
\bibinfo{author}{Alessandretti, L.}, \bibinfo{author}{Aslak, U.},
  \bibinfo{author}{Lehmann, S.}, \bibinfo{year}{2020}.
\newblock \bibinfo{title}{The scales of human mobility}.
\newblock \bibinfo{journal}{Nature} \bibinfo{volume}{587},
  \bibinfo{pages}{402--407}.
\bibitem[{Alves et~al.(2022)Alves, Pineda, Roster, Thielemann and
  Rodrigues}]{alves2022eeg}
\bibinfo{author}{Alves, C.L.}, \bibinfo{author}{Pineda, A.M.},
  \bibinfo{author}{Roster, K.}, \bibinfo{author}{Thielemann, C.},
  \bibinfo{author}{Rodrigues, F.A.}, \bibinfo{year}{2022}.
\newblock \bibinfo{title}{{EEG} functional connectivity and deep learning for
  automatic diagnosis of brain disorders: {A}lzheimer’s disease and
  schizophrenia}.
\newblock \bibinfo{journal}{Journal of Physics: Complexity}
  \bibinfo{volume}{3}, \bibinfo{pages}{025001}.
\bibitem[{Balcan et~al.(2009)Balcan, Colizza, Gon{\c{c}}alves, Hu, Ramasco and
  Vespignani}]{balcan2009multiscale}
\bibinfo{author}{Balcan, D.}, \bibinfo{author}{Colizza, V.},
  \bibinfo{author}{Gon{\c{c}}alves, B.}, \bibinfo{author}{Hu, H.},
  \bibinfo{author}{Ramasco, J.J.}, \bibinfo{author}{Vespignani, A.},
  \bibinfo{year}{2009}.
\newblock \bibinfo{title}{Multiscale mobility networks and the spatial
  spreading of infectious diseases}.
\newblock \bibinfo{journal}{Proceedings of the National Academy of Sciences}
  \bibinfo{volume}{106}, \bibinfo{pages}{21484--21489}.
\bibitem[{Barbosa et~al.(2018)Barbosa, Barthelemy, Ghoshal, James, Lenormand,
  Louail, Menezes, Ramasco, Simini and Tomasini}]{barbosa2018human}
\bibinfo{author}{Barbosa, H.}, \bibinfo{author}{Barthelemy, M.},
  \bibinfo{author}{Ghoshal, G.}, \bibinfo{author}{James, C.R.},
  \bibinfo{author}{Lenormand, M.}, \bibinfo{author}{Louail, T.},
  \bibinfo{author}{Menezes, R.}, \bibinfo{author}{Ramasco, J.J.},
  \bibinfo{author}{Simini, F.}, \bibinfo{author}{Tomasini, M.},
  \bibinfo{year}{2018}.
\newblock \bibinfo{title}{Human mobility: {M}odels and applications}.
\newblock \bibinfo{journal}{Physics Reports} \bibinfo{volume}{734},
  \bibinfo{pages}{1--74}.
\bibitem[{Bichara et~al.(2015)Bichara, Kang, Castillo-Chavez, Horan and
  Perrings}]{bichara2015sis}
\bibinfo{author}{Bichara, D.}, \bibinfo{author}{Kang, Y.},
  \bibinfo{author}{Castillo-Chavez, C.}, \bibinfo{author}{Horan, R.},
  \bibinfo{author}{Perrings, C.}, \bibinfo{year}{2015}.
\newblock \bibinfo{title}{{SIS} and {SIR} epidemic models under virtual
  dispersal}.
\newblock \bibinfo{journal}{Bulletin of Mathematical Biology}
  \bibinfo{volume}{77}, \bibinfo{pages}{2004--2034}.
\bibitem[{Brockmann(2009)}]{brockmann2009human}
\bibinfo{author}{Brockmann, D.}, \bibinfo{year}{2009}.
\newblock \bibinfo{title}{Human mobility and spatial disease dynamics}.
\newblock \bibinfo{journal}{Reviews of Nonlinear Dynamics and Complexity}
  \bibinfo{volume}{2}, \bibinfo{pages}{1--24}.
\bibitem[{Byrne et~al.(2020)Byrne, McEvoy, Collins, Hunt, Casey, Barber,
  Butler, Griffin, Lane, McAloon et~al.}]{byrne2020inferred}
\bibinfo{author}{Byrne, A.W.}, \bibinfo{author}{McEvoy, D.},
  \bibinfo{author}{Collins, A.B.}, \bibinfo{author}{Hunt, K.},
  \bibinfo{author}{Casey, M.}, \bibinfo{author}{Barber, A.},
  \bibinfo{author}{Butler, F.}, \bibinfo{author}{Griffin, J.},
  \bibinfo{author}{Lane, E.A.}, \bibinfo{author}{McAloon, C.}, et~al.,
  \bibinfo{year}{2020}.
\newblock \bibinfo{title}{Inferred duration of infectious period of
  {SARS-CoV-2}: rapid scoping review and analysis of available evidence for
  asymptomatic and symptomatic {COVID-19} cases}.
\newblock \bibinfo{journal}{BMJ Open} \bibinfo{volume}{10},
  \bibinfo{pages}{e039856}.
\bibitem[{Casadiego et~al.(2017)Casadiego, Nitzan, Hallerberg and
  Timme}]{casadiego2017model}
\bibinfo{author}{Casadiego, J.}, \bibinfo{author}{Nitzan, M.},
  \bibinfo{author}{Hallerberg, S.}, \bibinfo{author}{Timme, M.},
  \bibinfo{year}{2017}.
\newblock \bibinfo{title}{Model-free inference of direct network interactions
  from nonlinear collective dynamics}.
\newblock \bibinfo{journal}{Nature Communications} \bibinfo{volume}{8},
  \bibinfo{pages}{1--10}.
\bibitem[{Cavallaro and Dianin(2019)}]{cavallaro2019cross}
\bibinfo{author}{Cavallaro, F.}, \bibinfo{author}{Dianin, A.},
  \bibinfo{year}{2019}.
\newblock \bibinfo{title}{Cross-border commuting in {C}entral {E}urope:
  features, trends and policies}.
\newblock \bibinfo{journal}{Transport Policy} \bibinfo{volume}{78},
  \bibinfo{pages}{86--104}.
\bibitem[{Ching et~al.(2015)Ching, Lai and Leung}]{ching2015reconstructing}
\bibinfo{author}{Ching, E.S.}, \bibinfo{author}{Lai, P.Y.},
  \bibinfo{author}{Leung, C.}, \bibinfo{year}{2015}.
\newblock \bibinfo{title}{Reconstructing weighted networks from dynamics}.
\newblock \bibinfo{journal}{Physical Review E} \bibinfo{volume}{91},
  \bibinfo{pages}{030801}.
\bibitem[{Favier et~al.(2006)Favier, D{\'e}gallier, Rosa-Freitas, Boulanger,
  Costa~Lima, Luitgards-Moura, Menkes, Mondet, Oliveira, Weimann
  et~al.}]{favier2006early}
\bibinfo{author}{Favier, C.}, \bibinfo{author}{D{\'e}gallier, N.},
  \bibinfo{author}{Rosa-Freitas, M.G.}, \bibinfo{author}{Boulanger, J.P.},
  \bibinfo{author}{Costa~Lima, J.}, \bibinfo{author}{Luitgards-Moura, J.F.},
  \bibinfo{author}{Menkes, C.E.}, \bibinfo{author}{Mondet, B.},
  \bibinfo{author}{Oliveira, C.}, \bibinfo{author}{Weimann, E.}, et~al.,
  \bibinfo{year}{2006}.
\newblock \bibinfo{title}{Early determination of the reproductive number for
  vector-borne diseases: the case of dengue in {B}razil}.
\newblock \bibinfo{journal}{Tropical Medicine \& International Health}
  \bibinfo{volume}{11}, \bibinfo{pages}{332--340}.
\bibitem[{Findlater and Bogoch(2018)}]{findlater2018human}
\bibinfo{author}{Findlater, A.}, \bibinfo{author}{Bogoch, I.I.},
  \bibinfo{year}{2018}.
\newblock \bibinfo{title}{Human mobility and the global spread of infectious
  diseases: a focus on air travel}.
\newblock \bibinfo{journal}{Trends in Parasitology} \bibinfo{volume}{34},
  \bibinfo{pages}{772--783}.
\bibitem[{Gao and Yan(2022)}]{gao2022autonomous}
\bibinfo{author}{Gao, T.T.}, \bibinfo{author}{Yan, G.}, \bibinfo{year}{2022}.
\newblock \bibinfo{title}{Autonomous inference of complex network dynamics from
  incomplete and noisy data}.
\newblock \bibinfo{journal}{Nature Computational Science} \bibinfo{volume}{2},
  \bibinfo{pages}{160--168}.
\bibitem[{Gillespie(2001)}]{gillespie2001approximate}
\bibinfo{author}{Gillespie, D.T.}, \bibinfo{year}{2001}.
\newblock \bibinfo{title}{Approximate accelerated stochastic simulation of
  chemically reacting systems}.
\newblock \bibinfo{journal}{The Journal of Chemical Physics}
  \bibinfo{volume}{115}, \bibinfo{pages}{1716--1733}.
\bibitem[{Gillespie and Petzold(2003)}]{gillespie2003improved}
\bibinfo{author}{Gillespie, D.T.}, \bibinfo{author}{Petzold, L.R.},
  \bibinfo{year}{2003}.
\newblock \bibinfo{title}{Improved leap-size selection for accelerated
  stochastic simulation}.
\newblock \bibinfo{journal}{The Journal of Chemical Physics}
  \bibinfo{volume}{119}, \bibinfo{pages}{8229--8234}.
\bibitem[{Hempel and Earn(2020)}]{hempel2020estimating}
\bibinfo{author}{Hempel, K.}, \bibinfo{author}{Earn, D.J.},
  \bibinfo{year}{2020}.
\newblock \bibinfo{title}{Estimating epidemic coupling between populations from
  the time to invasion}.
\newblock \bibinfo{journal}{Journal of the Royal Society Interface}
  \bibinfo{volume}{17}, \bibinfo{pages}{20200523}.
\bibitem[{Jandarov et~al.(2014)Jandarov, Haran, Bj{\o}rnstad and
  Grenfell}]{jandarov2014emulating}
\bibinfo{author}{Jandarov, R.}, \bibinfo{author}{Haran, M.},
  \bibinfo{author}{Bj{\o}rnstad, O.}, \bibinfo{author}{Grenfell, B.},
  \bibinfo{year}{2014}.
\newblock \bibinfo{title}{Emulating a gravity model to infer the spatiotemporal
  dynamics of an infectious disease}.
\newblock \bibinfo{journal}{Journal of the Royal Statistical Society: Series C
  (Applied Statistics)} \bibinfo{volume}{63}, \bibinfo{pages}{423--444}.
\bibitem[{Ma(2020)}]{ma2020estimating}
\bibinfo{author}{Ma, J.}, \bibinfo{year}{2020}.
\newblock \bibinfo{title}{Estimating epidemic exponential growth rate and basic
  reproduction number}.
\newblock \bibinfo{journal}{Infectious Disease Modelling} \bibinfo{volume}{5},
  \bibinfo{pages}{129--141}.
\bibitem[{Mari et~al.(2012)Mari, Bertuzzo, Righetto, Casagrandi, Gatto,
  Rodriguez-Iturbe and Rinaldo}]{mari2012modelling}
\bibinfo{author}{Mari, L.}, \bibinfo{author}{Bertuzzo, E.},
  \bibinfo{author}{Righetto, L.}, \bibinfo{author}{Casagrandi, R.},
  \bibinfo{author}{Gatto, M.}, \bibinfo{author}{Rodriguez-Iturbe, I.},
  \bibinfo{author}{Rinaldo, A.}, \bibinfo{year}{2012}.
\newblock \bibinfo{title}{Modelling cholera epidemics: the role of waterways,
  human mobility and sanitation}.
\newblock \bibinfo{journal}{Journal of The Royal Society Interface}
  \bibinfo{volume}{9}, \bibinfo{pages}{376--388}.
\bibitem[{Mertel et~al.(2022)Mertel, Vysko{\v{c}}il, Sch{\"u}ler,
  Schlechte-We{\l}nicz and Calabrese}]{mertel2022fine}
\bibinfo{author}{Mertel, A.}, \bibinfo{author}{Vysko{\v{c}}il, J.},
  \bibinfo{author}{Sch{\"u}ler, L.}, \bibinfo{author}{Schlechte-We{\l}nicz,
  W.}, \bibinfo{author}{Calabrese, J.M.}, \bibinfo{year}{2022}.
\newblock \bibinfo{title}{Fine-scale variation in the effect of national border
  on {COVID-19} spread: {A} case study of the {S}axon-{C}zech border region}.
\newblock \bibinfo{journal}{medRxiv} .
\bibitem[{Pinho et~al.(2010)Pinho, Ferreira, Esteva, Barreto, Morato~e Silva
  and Teixeira}]{pinho2010modelling}
\bibinfo{author}{Pinho, S.T.R.d.}, \bibinfo{author}{Ferreira, C.P.},
  \bibinfo{author}{Esteva, L.}, \bibinfo{author}{Barreto, F.R.},
  \bibinfo{author}{Morato~e Silva, V.}, \bibinfo{author}{Teixeira, M.},
  \bibinfo{year}{2010}.
\newblock \bibinfo{title}{Modelling the dynamics of dengue real epidemics}.
\newblock \bibinfo{journal}{Philosophical Transactions of the Royal Society A:
  Mathematical, Physical and Engineering Sciences} \bibinfo{volume}{368},
  \bibinfo{pages}{5679--5693}.
\bibitem[{Schl{\"a}pfer et~al.(2021)Schl{\"a}pfer, Dong, O’Keeffe, Santi,
  Szell, Salat, Anklesaria, Vazifeh, Ratti and West}]{schlapfer2021universal}
\bibinfo{author}{Schl{\"a}pfer, M.}, \bibinfo{author}{Dong, L.},
  \bibinfo{author}{O’Keeffe, K.}, \bibinfo{author}{Santi, P.},
  \bibinfo{author}{Szell, M.}, \bibinfo{author}{Salat, H.},
  \bibinfo{author}{Anklesaria, S.}, \bibinfo{author}{Vazifeh, M.},
  \bibinfo{author}{Ratti, C.}, \bibinfo{author}{West, G.B.},
  \bibinfo{year}{2021}.
\newblock \bibinfo{title}{The universal visitation law of human mobility}.
\newblock \bibinfo{journal}{Nature} \bibinfo{volume}{593},
  \bibinfo{pages}{522--527}.
\bibitem[{Senapati et~al.(2019)Senapati, Sardar, Ganguly, Ganguly,
  Chattopadhyay and Chattopadhyay}]{senapati2019impact}
\bibinfo{author}{Senapati, A.}, \bibinfo{author}{Sardar, T.},
  \bibinfo{author}{Ganguly, K.S.}, \bibinfo{author}{Ganguly, K.S.},
  \bibinfo{author}{Chattopadhyay, A.K.}, \bibinfo{author}{Chattopadhyay, J.},
  \bibinfo{year}{2019}.
\newblock \bibinfo{title}{Impact of adult mosquito control on dengue prevalence
  in a multi-patch setting: a case study in {K}olkata (2014--2015)}.
\newblock \bibinfo{journal}{Journal of Theoretical Biology}
  \bibinfo{volume}{478}, \bibinfo{pages}{139--152}.
\bibitem[{Shandilya and Timme(2011)}]{shandilya2011inferring}
\bibinfo{author}{Shandilya, S.G.}, \bibinfo{author}{Timme, M.},
  \bibinfo{year}{2011}.
\newblock \bibinfo{title}{Inferring network topology from complex dynamics}.
\newblock \bibinfo{journal}{New Journal of Physics} \bibinfo{volume}{13},
  \bibinfo{pages}{013004}.
\bibitem[{Shi et~al.(2021)Shi, Shen, Jin, Shi, Wang and
  Boccaletti}]{shi2021inferring}
\bibinfo{author}{Shi, L.}, \bibinfo{author}{Shen, C.}, \bibinfo{author}{Jin,
  L.}, \bibinfo{author}{Shi, Q.}, \bibinfo{author}{Wang, Z.},
  \bibinfo{author}{Boccaletti, S.}, \bibinfo{year}{2021}.
\newblock \bibinfo{title}{Inferring network structures via signal lasso}.
\newblock \bibinfo{journal}{Physical Review Research} \bibinfo{volume}{3},
  \bibinfo{pages}{043210}.
\bibitem[{Simini et~al.(2012)Simini, Gonz{\'a}lez, Maritan and
  Barab{\'a}si}]{simini2012universal}
\bibinfo{author}{Simini, F.}, \bibinfo{author}{Gonz{\'a}lez, M.C.},
  \bibinfo{author}{Maritan, A.}, \bibinfo{author}{Barab{\'a}si, A.L.},
  \bibinfo{year}{2012}.
\newblock \bibinfo{title}{A universal model for mobility and migration
  patterns}.
\newblock \bibinfo{journal}{Nature} \bibinfo{volume}{484},
  \bibinfo{pages}{96--100}.
\bibitem[{Sysoeva et~al.(2021)Sysoeva, Sysoev, Prokhorov, Ponomarenko and
  Bezruchko}]{sysoeva2021reconstruction}
\bibinfo{author}{Sysoeva, M.V.}, \bibinfo{author}{Sysoev, I.V.},
  \bibinfo{author}{Prokhorov, M.D.}, \bibinfo{author}{Ponomarenko, V.I.},
  \bibinfo{author}{Bezruchko, B.P.}, \bibinfo{year}{2021}.
\newblock \bibinfo{title}{Reconstruction of coupling structure in network of
  neuron-like oscillators based on a phase-locked loop}.
\newblock \bibinfo{journal}{Chaos, Solitons \& Fractals} \bibinfo{volume}{142},
  \bibinfo{pages}{110513}.
\bibitem[{Timme(2007)}]{timme2007revealing}
\bibinfo{author}{Timme, M.}, \bibinfo{year}{2007}.
\newblock \bibinfo{title}{Revealing network connectivity from response
  dynamics}.
\newblock \bibinfo{journal}{Physical Review Letters} \bibinfo{volume}{98},
  \bibinfo{pages}{224101}.
\bibitem[{Wilder-Smith and Gubler(2008)}]{wilder2008geographic}
\bibinfo{author}{Wilder-Smith, A.}, \bibinfo{author}{Gubler, D.J.},
  \bibinfo{year}{2008}.
\newblock \bibinfo{title}{Geographic expansion of dengue: the impact of
  international travel}.
\newblock \bibinfo{journal}{Medical Clinics of North America}
  \bibinfo{volume}{92}, \bibinfo{pages}{1377--1390}.
\bibitem[{Xia et~al.(2004)Xia, Bj{\o}rnstad and Grenfell}]{xia2004measles}
\bibinfo{author}{Xia, Y.}, \bibinfo{author}{Bj{\o}rnstad, O.N.},
  \bibinfo{author}{Grenfell, B.T.}, \bibinfo{year}{2004}.
\newblock \bibinfo{title}{Measles metapopulation dynamics: a gravity model for
  epidemiological coupling and dynamics}.
\newblock \bibinfo{journal}{The American Naturalist} \bibinfo{volume}{164},
  \bibinfo{pages}{267--281}.
\bibitem[{Yang et~al.(2015)Yang, Zhang, Kargbo, Yang, Chen, Chen, Kamara,
  Kargbo, Kandula, Karspeck et~al.}]{yang2015transmission}
\bibinfo{author}{Yang, W.}, \bibinfo{author}{Zhang, W.},
  \bibinfo{author}{Kargbo, D.}, \bibinfo{author}{Yang, R.},
  \bibinfo{author}{Chen, Y.}, \bibinfo{author}{Chen, Z.},
  \bibinfo{author}{Kamara, A.}, \bibinfo{author}{Kargbo, B.},
  \bibinfo{author}{Kandula, S.}, \bibinfo{author}{Karspeck, A.}, et~al.,
  \bibinfo{year}{2015}.
\newblock \bibinfo{title}{Transmission network of the 2014--2015 {E}bola
  epidemic in {Sierra Leone}}.
\newblock \bibinfo{journal}{Journal of The Royal Society Interface}
  \bibinfo{volume}{12}, \bibinfo{pages}{20150536}.

\end{thebibliography}

\end{document}